\documentclass{jcst}

\title{Optimization of a Line Detection Algorithm for Autonomous Vehicles on a RISC-V with Accelerator\\
\large Optimización de un Algoritmo de Detección de Líneas para Vehículos Autónomos en un RISC-V con Acelerador}
\author[1\orcid{0000-0002-2870-7679}]{María José Belda}
\author[1\orcid{0000-0002-1821-124X}]{Katzalin Olcoz}
\author[1\orcid{0000-0002-2773-3023}]{Fernando Castro}
\author[1\orcid{0000-0003-0974-2687}]{Francisco Tirado}

\affil[1]{Complutense University of Madrid, Madrid 28040, España \authorcr
\{mbelda,katzalin,fcastror,ptirado\}@ucm.es }

\usepackage{xcolor}

\usepackage{siunitx}

\begin{document}

\maketitle

\begin{abstract}
In recent years, autonomous vehicles have attracted the attention of many research groups, both in academia and business, including researchers from leading companies such as Google, Uber and Tesla. This type of vehicles are equipped with systems that are subject to very strict requirements, essentially aimed at performing safe operations --both for potential passengers and pedestrians-- as well as carrying out the processing needed for decision making in real time. In many instances, general-purpose processors alone cannot ensure that these safety, reliability and real-time requirements are met, so it is common to implement heterogeneous systems by including accelerators. \textcolor{black}{This paper explores the acceleration of a line detection application in the autonomous car environment using a heterogeneous system consisting of a general-purpose RISC-V core and a domain-specific accelerator. In particular, the application is analyzed to identify the most computationally intensive parts of the code and it is adapted accordingly for more efficient processing.} Furthermore, the code is executed on the aforementioned hardware platform to verify that the execution effectively meets the existing requirements in autonomous vehicles, experiencing a \textcolor{black}{3.7x} speedup with respect to running without accelerator.
\end{abstract}


\renewcommand{\abstractname}{Resumen}
\begin{abstract}
En los últimos años los vehículos autónomos están atrayendo la atención de muchos grupos de investigación, tanto del ámbito académico como del empresarial, entre los que se incluyen investigadores pertenecientes a empresas punteras como Google, Uber o Tesla. Los sistemas de los que están dotados este tipo de vehículos están sometidos a requisitos muy estrictos relacionados esencialmente con la realización de operaciones seguras, tanto para los potenciales pasajeros como para los peatones, así como con que el procesamiento necesario para la toma de decisiones se realice en tiempo real. En muchas ocasiones, los procesadores de propósito general no pueden por sí solos garantizar el cumplimiento de estos requisitos de seguridad, fiabilidad y tiempo real, por lo que es común implementar sistemas heterogéneos mediante la inclusión de aceleradores. En este artículo se explora la aceleración de una aplicación de detección de líneas en el entorno de vehículos autónomos utilizando para ello un sistema heterogéneo formado por un core RISC-V de propósito general y un acelerador de dominio específico. En particular, se analiza dicha aplicación para identificar las partes del código más costosas computacionalmente y se adapta el código para un procesamiento más eficiente. Además, se ejecuta dicho código en la mencionada plataforma hardware y se comprueba que su procesamiento efectivamente cumple con los requisitos presentes en los vehículos autónomos, experimentando una reducción de 3.7x en su tiempo de ejecución con respecto a su ejecución sin acelerador. 
\end{abstract}

\palabrasclaves{Vehículos autónomos, Firesim, Procesamiento de imágenes, Acelerador de matrices,  RISC-V}

\section{Introduction}

In the technological era in which we live, we every day strive to make all the usual tasks as automatic as possible in order to gain free time. In addition, we try to achieve scenarios that are impossible right now, such as smarter power grids, fully autonomous vehicles or smart cities. This is why the Internet of Things (IoT) arises, as we need new technologies to design these systems. Most of them are on-board systems, so they need to get a trade-off between power consumption and \textcolor{black}{delivered performance}. In particular, \textcolor{black}{in this work we focus on} autonomous vehicles.

Autonomous driving systems aim to enable vehicles to drive on the road without human intervention~\cite{asplos_18,ieee_micro_15,bose_21}. Therefore, these systems must guarantee the safety and integrity of the vehicle, for which they must take a series of decisions in real time, including moving the steering wheel to ensure that the correct trajectory is followed, detecting obstacles in the path (pedestrians, animals, objects...), activating the braking mechanism when necessary and others. For this purpose, it is essential that the vehicle has a camera that records images of the route and processes them in real time to ensure the correct and safe operation of the vehicle. This image processing requires considerable computing power, but at the same time, when talking about on-board systems, it is essential to keep energy consumption at low levels so the vehicle does not loose autonomy~\cite{autonomia-va}. 

For these reasons, autonomous vehicles require \textcolor{black}{on-board automatic systems to process the recorded images that allow certain operations such as line and edge detection. Currently, the most widely used algorithms for this type of processing require high performance and their basic kernel is matrix and vector multiplication. It is therefore highly desirable that this type of algorithms could be executed in one of the many domain specific accelerators that have emerged in recent years.} 

\textcolor{black}{In this paper we propose to accelerate a line detection application employed in autonomous cars by using different heterogeneous systems made up of a general-purpose RISC-V core working at low frequency and a domain-specific accelerator. For this purpose, the application is deeply analyzed in order to identify the computationally intensive parts of the code and  adapted consequently for a more efficient processing}. \textcolor{black}{As it will be explained} in Section 3, the hardware platform used in this work includes, on the one hand, a general-purpose BOOM processor, which is an out-of-order RISC-V core~\cite{boom}, and on the other hand, the Gemmini~\cite{gemmini-dac21} accelerator, specifically designed for matrix multiplication. This platform was chosen because the RISC-V architecture, in addition to being open source, allows the integration of accelerators and their potential adaptation in a very simple way. Furthermore, the RISC-V instruction set architecture (ISA) is highly modular, allowing to choose exactly the functionalities needed, which is especially useful in IoT environments.

This paper leverages two image processing algorithms: 1) the Canny algorithm for edge detection of an image, and 2) the Hough transform, oriented to find imperfect instances of objects within a certain class of shapes by means of a voting procedure. In Section 4 we perform a detailed analysis of both algorithms codes, in order to identify the computational load of the different functions included in these programs, as well as the available parallelism. Moreover, we schedule some functions to run on the accelerator, while the rest of the algorithm is executed on the processor, aimed to optimize the total execution time and consequently to meet the strict requirements of performance, consumption and safety imposed by autonomous vehicles. The experimental evaluation carried out in Section 5  reports a speedup of \textcolor{black}{3.7x} when executing these algorithms with respect to the baseline where no accelerator is employed. Finally, Section 6 concludes the paper.

\section{Basic notions and state of the art}

In this section we explain some basic notions related to autonomous vehicles. We also provide details on the RISC-V-based development environment that we employ, including the tools used that make it possible the evaluation of the proposal presented in this paper.

\subsection{Autonomous vehicles}

\begin{figure*}[ht]
\centering
\includegraphics[width=0.85\textwidth]{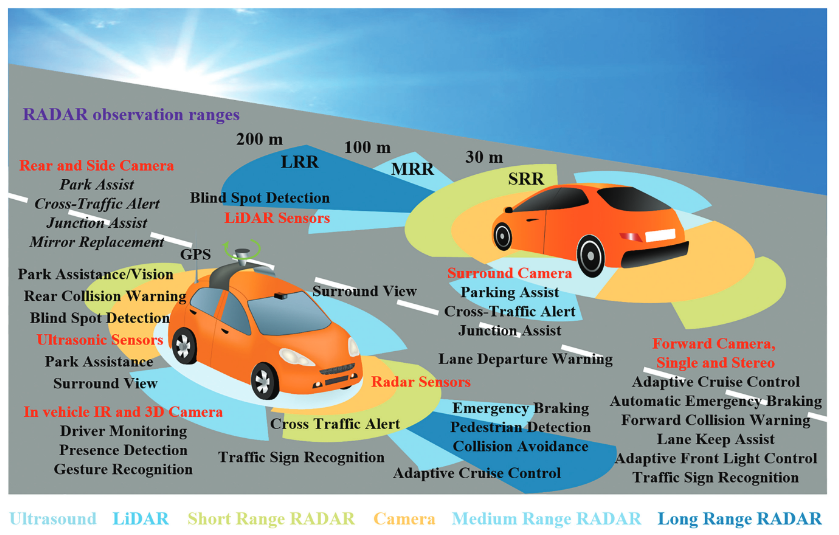}
\caption{\label{fig:sensores_va}Integrated sensors on an autonomous vehicle~\cite{sensors-av}.}
\end{figure*}

Autonomous vehicles are equipped with several sensors, as shown in Fig.~\ref{fig:sensores_va}, including video cameras, which are responsible for obtaining the data that serve as input to the processing system. The purpose of this data processing is to recognize the environment which the vehicle is driving through, and as a result, to make the appropriate decisions at any time, so as to ensure that the vehicle can reach its destination efficiently and safely. In this aspect, autonomous vehicles have levels of driving automation from 0 (No automation) to 5 (Full automation), as explained in~\cite{levels-av}.
In the first levels, from 0 to 2, the vehicle has very little capacity to act (in level 2 it can only perform steering and acceleration) and all the responsibility lies on the driver. In contrast, the automatic system monitors the driving environment in levels 3 to 5, being this last one the ideal scenario in which the vehicle is completely autonomous, even not providing controls for the driver. So, there is a gap between levels 2 and 3. Between these levels there is also a technological gap, since generating hardware and software capable of monitoring the environment in real time becomes significantly difficult. However, this gap is progressively disappearing and this work aims to contribute to this.

Notably, certain safety decisions are related to the correct recognition of the trajectory to be followed by the vehicle, based on the images recorded by the camera. In addition to allowing the car to follow the correct route, this functionality also involves restricting the likelihood of an accident. For this purpose, computer vision algorithms are commonly used in these processing systems~\cite{coppola2016connected,rateke2019passive} and, in particular, quite approaches use Canny algorithm to detect edges combined with the Hough transform to detect road lines~\cite{stateofart_linedetection_1,stateofart_linedetection_2}. Therefore, in this paper we focus on improving the performance of these algorithms which are the basis of lane detection. The problem with these algorithms is their very high computational cost. \textcolor{black}{In addition to this, there is a need for data processing to be performed in real time so that the vehicle could react with immediacy to changing situations that may occur during the journey.} It is also highly desirable that the energy consumption associated with such processing was as low as possible, so that the vehicle's autonomy was not affected.

\begin{figure}[ht]
\centering
\includegraphics[width=0.45\textwidth]{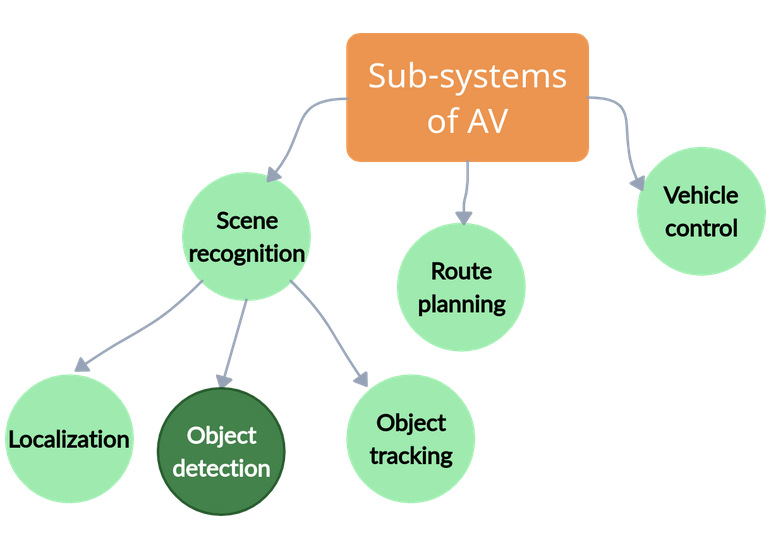}
\caption{\label{fig:av_subsystem}\textcolor{black}{Schematic of the subsystems of an autonomous vehicle.}}
\end{figure}

Autonomous driving systems are essentially composed of three classes of sub-systems~\cite{asplos_18,ieee_micro_15}: \emph{scene recognition}, \emph{route planning} and \emph{vehicle control}, consisting of a set of algorithms each. In particular, \textcolor{black}{as shown in Fig.~\ref{fig:av_subsystem},} \emph{scene recognition}, the class in which this article falls, comprises three essential tasks, namely 1) \emph{localization}, which precisely establishes the vehicle's location, 2) \emph{object detection}, which identifies objects of interest in the vehicle's environment (such as other vehicles, pedestrians or road signs, with the aforementioned objective of avoiding accidents and also traffic violations), and 3) \emph{object tracking}, which, since the object detection algorithm is carried out on each frame of the image, is responsible for relating its results to other frames in order to predict the trajectories of moving objects. These three tasks account for a very high percentage of the total computation time required~\cite{asplos_18} and therefore constitute bottlenecks that significantly limit the ability of conventional processors to satisfy the existing restrictions in the design of this type of systems. For this reason, it is being proposed to incorporate some type of accelerator to the on-board processing systems that helps the processor to fulfill the strict time limits in which it must operate.
 
\subsection{RISC-V-based development environment}

In order to carry out the implementation and evaluation of our proposal, which will be explained in the following section, a series of software tools have been used, as detailed next:

\subsubsection{Chipyard.}

Chipyard~\cite{chipyard} is an environment for the design and evaluation of hardware systems that consists of a set of tools and libraries designed to provide an integration path between open-source tools and commercial tools for the development of Systems on Chip (SoC). The environment provides a range of components for design construction as well as for compilation and simulation. Among these components there are several RISC-V cores and accelerators, including the BOOM core and Gemmini accelerator that make up the heterogeneous system chosen in this paper and that will be detailed in Section \ref{sec:hw}. The simulation of the complete system accelerated with FPGA is one of the types of simulation supported by Chipyard, using the FireSim tool described below.

\subsubsection{FireSim.}

FireSim~\cite{firesim} is a hardware simulation platform that runs on Amazon cloud services and automatically deploys the FPGA services in the cloud when needed. In particular, the user can generate the RTL of an own design and run it on these FPGAs, obtaining the same results as if the circuit was physically deployed.

\subsubsection{Amazon Web Services.}

Amazon Web Services~\cite{aws} is a cloud services platform that offers from training courses in new technologies --such as artificial intelligence or IoT-- to infrastructure services --such as storage or cloud computing. We focus on cloud computing because it offers a wide range of hardware platforms, including EC2 F1 instances that correspond to FPGAs, giving us the versatility we need to synthesize designs and to simulate the execution of applications on them.

\section{Platform design}\label{sec:hw}

\begin{figure}[ht]
\centering
\includegraphics[width=0.45\textwidth]{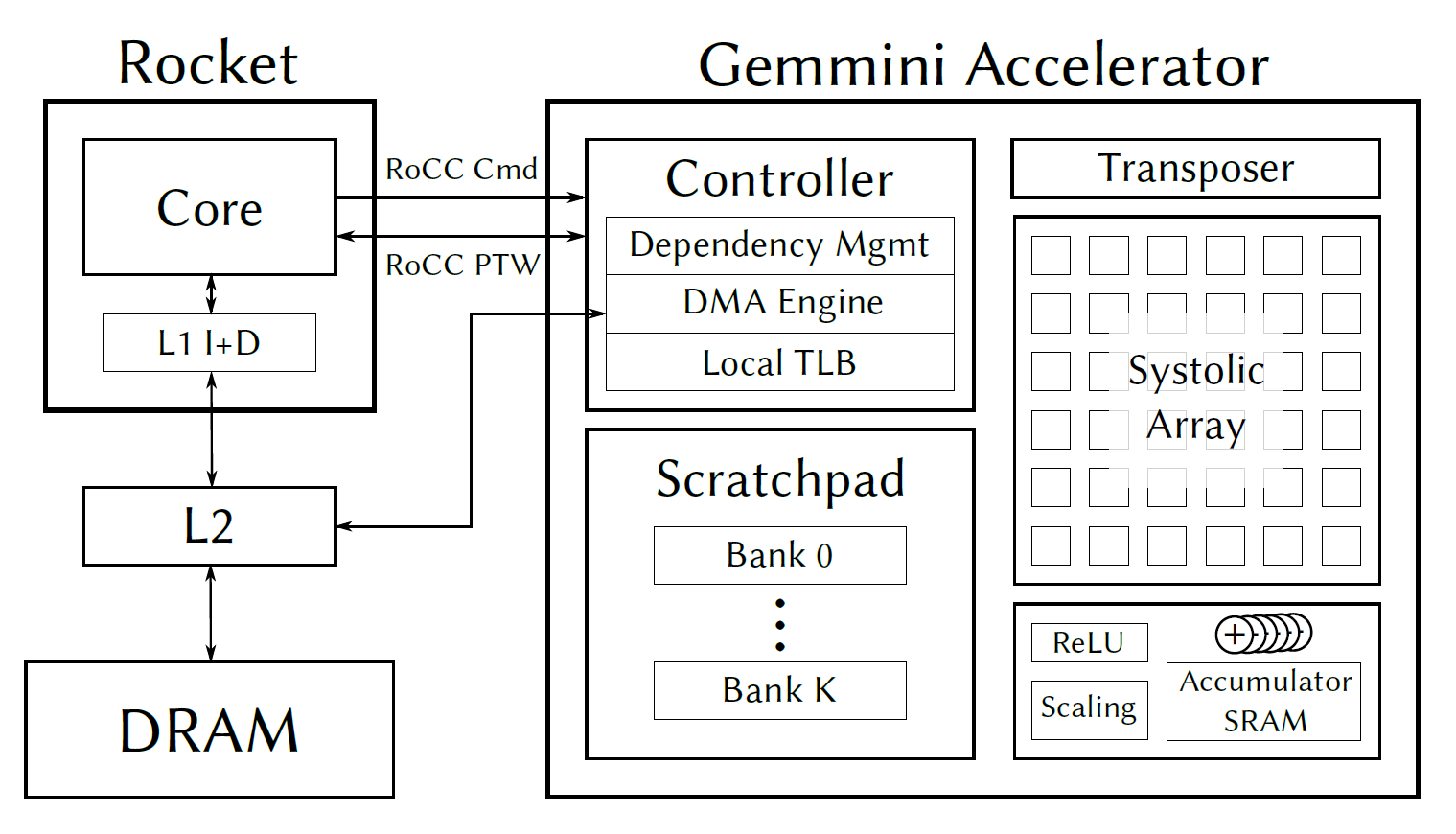}
\caption{Architecture of our heterogeneous platform~\protect\cite{gemmini-dac21}.}
\label{fig:gemmini}
\end{figure}

The platform employed in our experiments features a general-purpose processor equipped with an accelerator --implemented as a systolic array architecture-- for matrix multiplication. Both components have been developed by the Computer Architecture group at Berkeley University~\cite{gemmini-dac21}. The accelerator communicates with the processor through the RoCC (Rocket Co-Processor) interface, which allows the accelerator to receive the specific instructions that the processor sends, as shown in Fig.~\ref{fig:gemmini}. In the following two sections we describe the processors and the accelerator used.

\subsection{Processors}

As Fig.~\ref{fig:gemmini} illustrates, our system features a core plus an accelerator. In our experiments we opted to employ either the Rocket or the BOOM (Berkeley Out-of-Order Machine) processor. Both of them are written in Chisel and implement the RV64GC instruction set. Also, they are easily parameterizable and can be synthesized. Notably, the cores are configured by using the Rocket Chip SoC generator~\cite{Asanovic}.

The main differences between both cores lie in the pipeline characteristics: while the Rocket core features an in-order 5-stage pipeline, the BOOM core is equipped with a deeper out-of-order pipeline, which is inspired by those of MIPS R10000 and Alpha 212645~\cite{boom}. Consequently, the BOOM core is expected to deliver higher performance when executing our line detection algorithm. However, this comes at the expense of higher energy consumption than that of the Rocket core. Therefore, we experiment with both processors in order to check if the speedup reported by the BOOM core is significant enough to cancel out the energy constraints.

\subsection{The Gemmini Accelerator}
The Gemmini matrix multiplication accelerator relies on a 2D systolic array architecture, as shown in Fig.~\ref{fig:gemmini}, to perform matrix multiplications in an efficient fashion. In addition to this systolic array, it also features a scratchpad memory with multiple banks and an accumulator, which has more bits than that of the systolic array. Besides, the implementation allows to choose, at compile time, between two specific calculation mechanisms: output-stationary or weight-stationary.

 Customized instructions --out of RISC-V standard-- are available for the Gemmini accelerator, so that it is equipped with its own instruction queues that make it possible to execute concurrently with the processor. The Gemmini programming model can be broken down into three different levels. In the high-level we can run Open Neural Network Exchange (ONNX) models, being the accelerator itself in charge of mapping the ONNX kernel to the accelerator by means of dynamic dispatch. In the mid-level we use a hand-tuned library including C macros to perform data transfers between the main memory and the accelerator's scratchpad memory, which should be explicitly defined, as well as to automate the calculation of the block size used to split a matrix and to perform the full multiplication in a transparent way for users. Among available functions we highlight the following: \emph{tiled\_matmul}, to run a tiled matrix multiplication with hardcoded tiling factors; \emph{tiled\_conv}, to apply a convolution with hardcoded tiling factors; \emph{tiled\_matmul\_auto}, to run a tiled matrix multiplication with automatically calculated tiling factors; \emph{gemmini\_mvin}, to move data from the main memory to the scratchpad and \emph{gemmini\_mvout}, to move data from the scratchpad to the main memory. Finally, at the low-level, we can write our own mid-level kernels with low-level assembly instructions.

\section{Adapting image processing algorithms}
\label{cap:algoritmosDeTratamientodeImagenes}

As stated previously, the aim of this work is to accelerate image processing algorithms employed to guide autonomous vehicles. Notably, we focus on those algorithms targeted to detect road lines from road images. In this section we first introduce the basic algorithms used (the Canny algorithm and the Hough transform). Then, we show the full algorithm that we have employed in this work as starting point for line detection and, finally, we propose some changes to this algorithm oriented to improve its efficiency and performance without impacting on accuracy.

\subsection{Canny Algorithm}
Among the edge detection methods developed to date, the Canny algorithm is one of the methods more strictly defined that provides a satisfactory and reliable detection. Thus, it has become one of the most popular algorithms targeting edge detection.

This algorithm relies on calculus of variations, which allows to find an analytical function to approximate the real curve (i.e., the road lines) as accurately as possible. The procedure followed by the Canny algorithm~\cite{cannyAlgorithm} can be broken down into 5 stages as shown next:

\begin{enumerate}
    \item Noise reduction: applying the Gauss filter for image smoothing.
    \item To find the intensity gradient of the image.
    \item Magnitude threshold to the gradient: applying a threshold to the gradient for discarding edge false positives.
    \item Double threshold: applying again a threshold to the gradient for highlighting the potential edges.
    \item Hysteresis: removing weak or disconnected edges.
\end{enumerate}

Algorithm~\ref{alg:canny-code} shows the pseudo-code we employed to apply the Canny algorithm, broken down into the 5 stages aforementioned. Essentially, it includes multiplications of consecutive matrices and conditions checking in order to detect edge points.

\providecommand{\abs}[1]{\lvert#1\rvert}

\begin{algorithm}[ht]
\caption{Canny algorithm summarized pseudo-code.}\label{alg:canny-code}
\begin{algorithmic}[1]
\State float $NR \gets \text{mask} * \text{image}$ \Comment{Stage 1: Noise reduction}
\State float $G_x \gets \text{mask} * \text{NR}$ \Comment{Stage 2: Gradient intensity}
\State float $G_y \gets \text{mask} * \text{NR}$
\State float $G \gets \sqrt{G_x^2 + G_y^2} $
\State float $\phi \gets \arctan (\abs{G_y} / \abs{G_x})$
\If{$\phi[*] \geq \text{threshold}_{\phi}$} \Comment{Stage 3: Gradient threshold}
    \State float $\phi \in \{ 0, 45, 90, 135\}$
\EndIf
\If{$\phi[*] \geq \text{threshold}_{\phi} \ \&\& \ G[*] \geq \text{threshold}_G$} \Comment{Stage 4: Double threshold}
    \State int edge$[*] \gets 1$
\EndIf

\If{$G[*] \geq \text{threshold}_G \ \&\& \ \text{edge}[*] == 1$} \Comment{Stage 5: Hysteresis}
    \State int image\_out$[*] \gets 255$
\EndIf
    
\end{algorithmic}
\end{algorithm}

\subsection{Hough Transform}

The Hough transform is a technique of features extraction which is employed in multiple fields involving image processing, as computer vision or image digital processing. The goal of the algorithm is to find imperfect objects among certain classes of objects by means of a voting procedure. This procedure lies in creating a space with the values assigned to each pixel, so that the resulting local maximums in the so called accumulator space are the possible detected objects.

Generally, the classical Hough transform was only applied to the detection of straight lines, but in recent years it has been modified and currently it is employed for the detection of arbitrary curves, as ellipses or circles.

Algorithm~\ref{alg:hough-code} illustrates the code we employed to apply the Hough transform~\cite{houghTransform}. In this code, for each edge point previously detected with the Canny algorithm, the Hough transform draws a set of straight lines going through that point, recording the amount of lines going through each image pixel. Hence, those points with more lines going through them will correspond to a line in the original image.

\begin{algorithm}[ht]
\caption{Hough transform summarized pseudo-code.}\label{alg:hough-code}
\begin{algorithmic}[1]
\State For each edge point $(i,j)$
\If{image$[ i*width + j] \geq 250$}
\State $\theta \gets 0$
\While{$\theta \le 180$}
\State float $\rho \gets j * \cos{\theta} + i * \sin{\theta} $
\State accumulators$[(\rho + c_{\rho} )* 180 + \theta]$++
\State $\theta$++
\EndWhile
\EndIf
    
\end{algorithmic}
\end{algorithm}

\subsection{Line Detection}
Once we have described the two previous algorithms, we now employ a combination of both as well as another specific code targeted to detect with higher accuracy the lines that demarcate lanes in conventional ways.
For this purpose, given a certain input image, we first apply the Canny algorithm and then the Hough transform, so that we can apply a function (\emph{Get lines coordinates}) to detect lines in the resulting image. In Algorithm~\ref{alg:get-lines-code} we show the code of the mentioned function, which involves a search of local maximums in the preprocessed image and the generation of a straight line going through closest maximums.

\begin{algorithm}[ht]
\caption{Get lines coordinates algorithm summarized pseudo-code.}\label{alg:get-lines-code}
\begin{algorithmic}[1]
\State For each image point $(\rho, \theta)$
\If{accumulators$[*] \geq$ threshold} \Comment{It is a local maximum} 
\State $max \gets \text{accumulators[*]}$
\If{accumulators[neighbourhood(*)] $\ge max$} \Comment{We check its neighborhood}
\State $max \gets $ accumulators[neighbourhood(*)]
\EndIf
\EndIf

\State lines.add($x_1, y_1, x_2, y_2$) \Comment{We save the two points that demarcate the straight line}
    
\end{algorithmic}
\end{algorithm}

\subsection{Delivering higher performance}
\begin{figure*}[ht]
\centering
\includegraphics[width=0.85\textwidth]{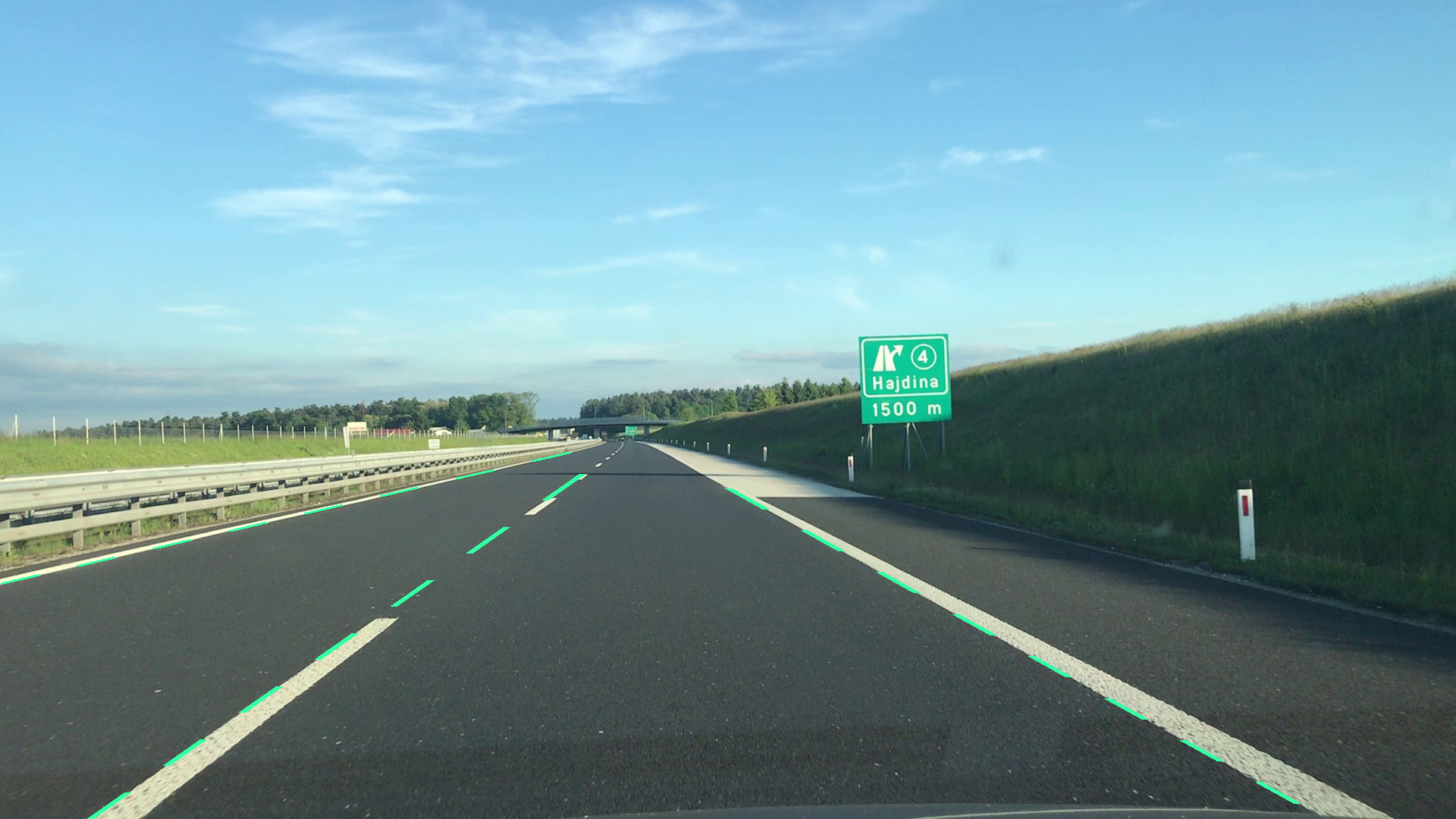}
\caption{Original image with detected lines highlighted in red.}
\label{cap3:fig:imagen_out}
\end{figure*}

In the previous sections we have described the original code of the algorithms, which indeed deploys many floating point variables. Therefore, it is advisable to replace them by integer variables without any loss in accuracy. We effectively made these changes in the code and we verified that no accuracy loss occurs when detecting lines in an image. Fig.~\ref{cap3:fig:imagen_out} shows the original image with detected lines highlighted in red. The analytical results corresponding to the lines detected with the original algorithm and with the simplified one do match, and also the second algorithm has performed slightly faster. Details on these modifications can be found in~\cite{TFM}.

Apart from this change, we also performed a profiling of the full code divided into three stages: 1) original image loading, 2) lines detection and 3) generation of an output image with the detected lines. Accordingly to the results obtained, we opted for not generating an output image (that is not needed by our system) due to the high cost associated, as shown in Table~\ref{cap3:table:profiling}. In doing so, we are able to reduce the execution time by \textcolor{black}{4.2x} as we can derive from data reported in Table~\ref{cap3:table:profiling-2}. 
It is worth noting that the time values illustrated in the mentioned tables are approximate since the profiling was not performed on the target platform, but on an Intel i7 processor running Linux. However, in order to report time values as accurate as possible, the algorithms were run several times so that the tables show the average values obtained. According to numbers from Table~\ref{cap3:table:profiling-2}, our attention is drawn to the line detection phase since it accounts for almost 70\% of the execution time.

\begin{table}[ht]
\caption{\label{cap3:table:profiling}Phased \textit{profiling} of the full code.}
\begin{tabular}{lcc}
                 & Time($\mu$s)             & \% over total \\ \hline
Image load       & 43803                    & 7,32\%                            \\ \hline
Line detection   & 98171                    & 16,42\%                           \\ \hline
Image generation & 456030                   & 76,26\%                           \\ \hline
Total            & 598004                   & \multicolumn{1}{l}{}  
\end{tabular}
\end{table}

\begin{table}[ht]
\caption{\label{cap3:table:profiling-2}Phased \textit{profiling} of the full code excluding the generation of the output image.}
\begin{tabular}{lcc}
               & Time($\mu$s)             & \% over total               \\ \hline
Image load     & 43485                    & \multicolumn{1}{c}{30,58\%} \\ \hline
Line detection & 98714                    & \multicolumn{1}{c}{69,42\%} \\ \hline
Total          & 142199                   &                            
\end{tabular}
\end{table}

In addition, we have performed another specific profiling of the stages of the line detection algorithm in order to know in which parts of the processing the acceleration efforts should be focused. Table~\ref{cap3:table:profiling-etapa2} illustrates that the most time-consuming part is the application of the Canny algorithm, which accounts for more than 87\% of the total execution time. Therefore, we will focus on accelerating this stage of image processing.

\begin{table}[ht]
\caption{\label{cap3:table:profiling-etapa2}Phased \textit{profiling} of the line detection algorithm.}
\begin{tabular}{lcc}
                & Time($\mu$s)   & \% over total        \\ \hline
Canny algorithm & 90265          & 87,64\%              \\ \hline
Hough transform & 12275          & 11,92\%              \\ \hline
Get coordinates & 459            & 0,45\%               \\ \hline
Total           & 102999         & \multicolumn{1}{l}{}
\end{tabular}
\end{table}

\section{Experimental results}
In this section, we first describe the hardware platforms as well as the workloads employed in our experiments, and then we detail the results obtained.

\subsection{Platforms generated}
\label{cap4:sect:arqGeneradas}
All the components used in the designs generated are written in Scala, so it is easy to modify their main features such as number of registers or number of Re-Order Buffer (ROB) entries.
Notably, we generate several designs: while all of them include one (or more) Rocket or BOOM cores, they may include or not the Gemmini accelerator.

Apart from the cores, for the sake of fairness the remaining components in the different designs generated (such as memory, clock frequency or buses) are the same in all of them. Hence, all designs have an L2 --shared in multicore platforms-- 4MB size. In order to optimize the design to fit into smaller FPGAs, the option MCRams is enabled in the FireSim platform configuration for all designs. This option allows the FPGA simulation tool (Golden Gate~\cite{goldengate}) to simulate the RAM via serialized accesses with a decoupled model~\cite{firesim}.

Platforms including the Gemmini accelerator can only be designed to work at 50MHz while the remaining ones can reach 80MHz. Thus, the later have been designed both at 50 and 80 MHz for a fair comparison against designs equipped with Gemmini. Notably, the platforms generated are: 

\begin{enumerate}
    \item \textit{Platform} 1: Rocket single core. \\
    This architecture includes a single \textit{Big} Rocket core.  There are four different sizes for the core, namely  \textit{Big, Medium, Small} and \textit{Tiny}, with different features such as the size of L1-cache. The \textit{Big} Rocket is the only one providing Floating Point Unit. It also has by default the parameters shown in Table \ref{cap4:table:platform_configs}. More information on the details of the configuration can be found in \cite{TFM}.

\begin{table*}[ht]
\centering
\caption{\label{cap4:table:platform_configs}Platform configuration options.}
\begin{tabular}{llcc}
\multicolumn{1}{c}{} & \multicolumn{1}{c}{} & Big Rocket & Large Boom \\ \hline
I\&D Cache           & Size                 & 16KB       & 32KB       \\
                     & Sets                 & 64         & 64         \\
                     & Ways                 & 4          & 8          \\
                     & Prefetching          & no         & disabled   \\ \hline
TLB                  & Sets                 & 1          & 1          \\
                     & Ways                 & 32         & 512        \\ \hline
BTB Entries          &                      & 28         & 28         \\ \hline
ROB Entries          &                      & no         & 96         \\ \hline
FPU                  &                      & yes        & yes        \\ \hline
Branch predictor \\
entries              &                      & no         & 128       
\end{tabular}
\end{table*}

    \item \textit{Platform} 2: Rocket dual core.
    
    This is the same configuration as Platform 1 but it includes two \textit{Big} Rocket cores. This dual configuration also has the option MTModels enabled in the FireSim platform configuration, so that each core is simulated with a separate thread of execution on a shared underlying physical implementation~\cite{firesim}.
    
    \item \textit{Platform} 3: Heterogeneous Rocket single core  + Gemmini Accelerator.
    
    This architecture is made up by a \textit{Big} Rocket core and a Gemmini matrix multiplication accelerator, which has been designed with default options: 16x16 8-bit systolic array, both dataflows supported (output-stationary and weight-stationary), float data type supported, a set of accumulator registers with 64B of total capacity, a 256KB scratchpad with 4 banks, a small TLB with 4 entries and a bus width of 128 bits.

    \item \textit{Platform} 4: BOOM Single core.
    
    This architecture includes a single \textit{Large} BOOM core. There are different macros for defining BOOM cores of \textit{Giga, Mega, Large, Medium} and \textit{Small} sizes. The main differences between the one that we are using and the rest is the number of entries in the ROB and some L1-cache parameters. Thus, in the configuration \textit{WithNLargeBooms} the value of notable parameters are shown in Table \ref{cap4:table:platform_configs}. More information on the details of the configuration can be found in \cite{TFM}. The Large size was chosen because it is just big enough to provide the required performance with minimum power consumption.

    \item \textit{Platform} 5: BOOM dual core.
    
    This is the same configuration as Platform 4 but it includes two \textit{Large} BOOM cores, with the MTModels option enabled.
    
    \item \textit{Platform} 6: Heterogeneous BOOM single core + Gemmini Accelerator.
    
     This architecture is made up by a \textit{Large} BOOM core and a Gemmini matrix multiplication accelerator, which has been designed with the default options explained earlier.
    
\end{enumerate}

\subsection{Workloads generated}
\label{cap4:sect:workloadsGenerados}
Different workloads were designed for running on the platforms described in the previous section. They are the following:
\begin{enumerate}
    
    \item \textit{Workload} 1: Multithreaded application on top of Linux buildroot distribution.
    \label{cap4:sect:workloadsGenerados:w1-multihilo}
    
    In this workload, a multithreaded application (with each thread computing the addition of 2 long arrays, as explained in \cite{TFM}) is executed on top of Linux. It has been specifically designed to fully exploit the parallel features of the platforms, so that it can be used to evaluate the maximum performance obtainable in the different multicore designs. This value will serve as an upper bound when we evaluate the performance achieved by our target application.
    
    \item \textit{Workload} 2: Line detection algorithm on top of Linux buildroot distribution.
    \label{cap4:sect:workloadsGenerados:w2-ldet-linux}
    
    In this workload, the modified version of the line detection application explained in Section \ref{cap:algoritmosDeTratamientodeImagenes} is executed on top of Linux. 
    
    \item \textit{Workload} 3: Line detection algorithm for bare-metal platforms with Gemmini.
    \label{cap4:sect:workloadsGenerados:w5-ldet-gemmini}
    
    In this workload, in addition to the modifications in Section 4, we have modified the line detection algorithm to add matrix multiplications. In the original version, this algorithm multiplies some mask values to a pixel neighborhood manually by writing the corresponding scalar multiplications. We have rewritten these multiplications in a matrix form, obtaining a 5x5 matrix for the mask and a 5x5 neighborhood matrix for each pixel. As for the platform, the differences with respect to the previous workload are that this platform includes a Gemmini accelerator for matrix multiplication and the fact that no operating system is available for this platform. Thus, matrix multiplications in the code have to be replaced by calls to a Gemmini multiplication. As previously explained, some C macros are provided with the designs that make it possible to easily programming the accelerator. First, data need to be moved from the main memory to the scratchpad memory in Gemmini, then the multiplication is performed in tiles and finally the results are transferred back to the main memory.
    We will use the \textit{tiled\_matmul\_auto} function that receives the dimensions of both matrices as input parameters and automatically splits the multiplication in blocks of suitable size for the systolic array and memory, thus performing the whole multiplication. Finally, system calls not available outside Linux were removed from the code and their functionality was implemented in an equivalent way.
    
\end{enumerate}

\subsection{Experiments}
In this section we show the results obtained from the execution of the workloads on the different platforms designed. The metrics measured are clock cycles and instructions retired provided by the performance counters of the target platforms.

\subsubsection{Experiment 1: Execution of a multithreaded application on single core and dual core platforms both with Rocket and BOOM cores.}
 The goal of this experiment is to verify the maximum performance attainable in the different platforms by using a massively parallel application. Therefore we employ \textit{Workload 1}, configured with as many independent threads as the number of cores in the system, i.e., 1 or 2 depending on the specific platform. 
 
 The target platforms in this case include both single and dual core processors (either Rocket or BOOM, running at 80MHz) that correspond to the \textit{Platforms} 1, 2, 4 and 5 previously described.
 
 The results of the experiment are shown in Table \ref{cap4:table:dualcoreN1_8}, both for a simulation in which the main loop is executed once (column labelled \textit{N\_times = 1}) and 8 times (column \textit{N\_times = 8}). 
 The number of clock cycles for the experiment with 8 iterations is 8 times the one of the single iteration experiment. 
 Besides, speedup of the dual core version with respect to the single core is very close to \textcolor{black}{2x} for both Rocket and BOOM. 
 Finally, comparing the performance of the different cores, BOOM achieves almost \textcolor{black}{2.2x} higher performance than Rocket, so that a single BOOM core outperforms a dual core Rocket running at the same frequency for this highly parallel application.

\begin{table}[ht]
\caption{Cycles when executing Workload 1 on Platforms 1, 2, 4 and 5.\label{cap4:table:dualcoreN1_8}}
\begin{tabular}{lll}
                              & N\_times = 1     & N\_times=8     \\ \cline{2-3} 
                              & \multicolumn{2}{c}{Cycles}        \\ \hline
Rocket singlecore             & \num{2.01e9}     & \num{1.59e10}  \\ \hline
BOOM singlecore               & \num{9.17e8}     & \num{7.31e9}   \\ \hline
Rocket dualcore               & \num{9.97e8}     & \num{7.99e9}   \\ \hline
BOOM dualcore                 & \num{4.53e8}     & \num{3.66e9}   \\ \hline
Speedup BOOM  \\ vs Rocket        & \textcolor{black}{2.19x}        & \textcolor{black}{2.18x}    \\ \hline
Speedup Rocket \\ dual vs single & \textcolor{black}{2.02x}        & \textcolor{black}{1.99x}    \\ \hline
Speedup BOOM \\ dual vs single   & \textcolor{black}{2.02x}        & \textcolor{black}{1.99x}    \\ \hline        
\end{tabular}
\end{table}

Thus, it has been verified that multithreaded applications are being correctly simulated in the multicore platforms, achieving the expected speedup.
Furthermore, the comparison between both types of cores has been established.

\subsubsection{Experiment 2: Execution of the line detection application on Rocket and BOOM single cores.}

This second experiment involves simulating the execution of the line detection application (\textit{workload} \ref{cap4:sect:workloadsGenerados:w2-ldet-linux}) on the Rocket and BOOM single core platforms employed in the previous experiment (Platforms 1 and 4), also running at 80MHz.
In Table~\ref{cap4:table:ldet-linux} we report the number of clock cycles and instructions retired corresponding to each of the different parts of the line detection algorithm, as well as the average cycles per instructions (CPI) value. \textcolor{black}{In addition, we calculate the actual time from the cycles and clock frequency, resulting in times of around half second. In particular, for the Rocket core we obtain a total execution time of 0.648s and for the Boom core 0.327s.} As shown, the CPI for the Hough transform is higher than 3 in both Rocket and BOOM platforms. Moreover, its execution on the BOOM processor almost matches the time reported on the Rocket platform, as the multiple data dependencies in the code make out-of-order capabilities useless.

\begin{table*}[]
\caption{Cycles, instructions retired and CPI when executing Workload 2 on Platforms 1 and 4 at 80MHz.\label{cap4:table:ldet-linux}}
\centering
\begin{tabular}{lllllr}
                       &             & Cycles        & Instructions     & CPI   
                                     & \textcolor{black}{Time(ms)}                           \\ \hline
Rocket singlecore      & Canny       & \num{2.18e9}  & \num{9.06e8}     & 2.40  
                                     & \textcolor{black}{648,38}                            \\ \cline{2-6} 
                       & Hough       & \num{3.32e8}  &  \num{9.35e7}    & 3.55  
                                     & \textcolor{black}{98,86}                             \\ \cline{2-6} 
                       & Coordinates & \num{6.49e6}  & \num{3.47e6}     &1.87   
                                     & \textcolor{black}{1,93}                               \\ \hline
Boom singlecore        & Canny       & \num{1.08e9}  & \num{9.06e8}     & 1.19  
                                     & \textcolor{black}{327,10}                             \\ \cline{2-6} 
                       & Hough       & \num{3.16e8}  & \num{9.35e7}     & 3.38  
                                     & \textcolor{black}{96,07}                             \\ \cline{2-6} 
                       & Coordinates & \num{3.2e6}   & \num{3.47e6}     & 0.92  
                                     & \textcolor{black}{0,97}                               \\ \hline
Speedup Boom vs Rocket & Canny       & \textcolor{black}{2.02x} & \textcolor{black}{1.00x}
                                     & \textcolor{black}{2.02x} & \textcolor{black}{1.98x}    \\ \cline{2-6} 
                       & Hough       & \textcolor{black}{1.05x} & \textcolor{black}{1.00x} 
                                     & \textcolor{black}{1.05x} & \textcolor{black}{1.03x}    \\ \cline{2-6} 
                       & Coordinates & \textcolor{black}{2.03x} & \textcolor{black}{1.00x} 
                                     & \textcolor{black}{2.03x} & \textcolor{black}{1.99x}
\end{tabular}
\end{table*}

On the other hand, the Canny and the GetCoordinates algorithms exhibit lower CPI numbers in both platforms, achieving a speedup of \textcolor{black}{2x} when executing on the Boom processor with respect to Rocket, due to the greater instruction level parallelism that can be extracted from both algorithms.
Recall that the Canny algorithm is the most relevant part of the line detection application, consuming close to 90\% of the total execution time (as shown in Table \ref{cap3:table:profiling-etapa2}). In conclusion, using the BOOM core for the execution of the workload is interesting in terms of the global speedup achieved.

\subsubsection{Experiment 3: Execution of the line detection application on heterogeneous platforms with a Rocket or BOOM single core and a Gemmini matrix multiplication accelerator.}
\label{cap4:sect:exp4-gemmini}

This experiment consists on simulating the execution of the modified line detection application (\textit{workload} \ref{cap4:sect:workloadsGenerados:w5-ldet-gemmini}) on the  heterogeneous single core platforms made up by a Rocket or BOOM processor plus a Gemmini matrix multiplication accelerator running at 50MHz.

\begin{table*}[ht]
\caption{Speedup results when executing Workload 2 on Platforms 1 and 4 at 80MHz, and Workload 3 on Platforms 3, 4, 6 at 50MHz, with respect to execution of Workload 3 on Platform 1 at 50 MHz.\label{cap4:table:ldet-gemmini}}
\centering
\begin{tabular}{lllll}
                        & \multicolumn{4}{l}{Speedup vs Rocket singlecore 50MHz} \\ \cline{2-5} 
                        & Canny      & Hough      & Coordinates      & Total     \\ \hline
Boom singlecore 50MHz   & \textcolor{black}{1.44x}          & \textcolor{black}{1.04x}          
                        & \textcolor{black}{1.85x}          & \textcolor{black}{1.41x}  \\ \hline
Rocket singlecore 80MHz & \textcolor{black}{2.26x}          & \textcolor{black}{0.98x}          
                        & \textcolor{black}{1.07x}          & \textcolor{black}{2.09x}  \\ \hline
Boom singlecore 80MHz   & \textcolor{black}{4.57x}          & \textcolor{black}{1.03x}
                        & \textcolor{black}{2.18x}          & \textcolor{black}{3.76x}  \\ \hline
Rocket + Gemmini 50MHz  & \textcolor{black}{2.54x}          & \textcolor{black}{1.16x}
                        & \textcolor{black}{1.03x}          & \textcolor{black}{2.36x}  \\ \hline
Boom + Gemmini 50MHz    & \textcolor{black}{4.43x}          & \textcolor{black}{1.07x}
                        & \textcolor{black}{1.98x}          & \textcolor{black}{3.70x}        
\end{tabular}
\end{table*}

Table \ref{cap4:table:ldet-gemmini} shows first the results obtained in the simulation of Workload 3 (line detection application for bare metal) on a Rocket single core (used as baseline for computing speedups) and a BOOM single core, both running at 50MHz. As the first row shows, BOOM is 41\% faster than Rocket. The execution results from the previous section, that is, those corresponding to Workload 2 (line detection application for Linux) on Rocket and BOOM single core at 80MHz are also compared to the baseline execution, achieving speedups of \textcolor{black}{2.09x} and \textcolor{black}{3.76x} respectively. It is worth noting that although the code of Workloads 2 and 3 does not exactly match, it performs the same functionality.
Finally, the results from the simulation of Workload 3 on heterogeneous platforms in which matrix multiplications are performed using the Gemmini accelerator are also recap in Table \ref{cap4:table:ldet-gemmini}. According to them, speedups of \textcolor{black}{2.36x} and \textcolor{black}{3.7x} are reported for Rocket and BOOM based platforms respectively, with respect to the baseline. Although these speedups can be considered as significant, they are far from the maximum values attainable by the accelerator. The reason is that the size of the matrices employed is smaller than that of the systolic array, which indeed is not fully utilized.

\begin{figure*}[ht]
\centering
\includegraphics[width=\textwidth]{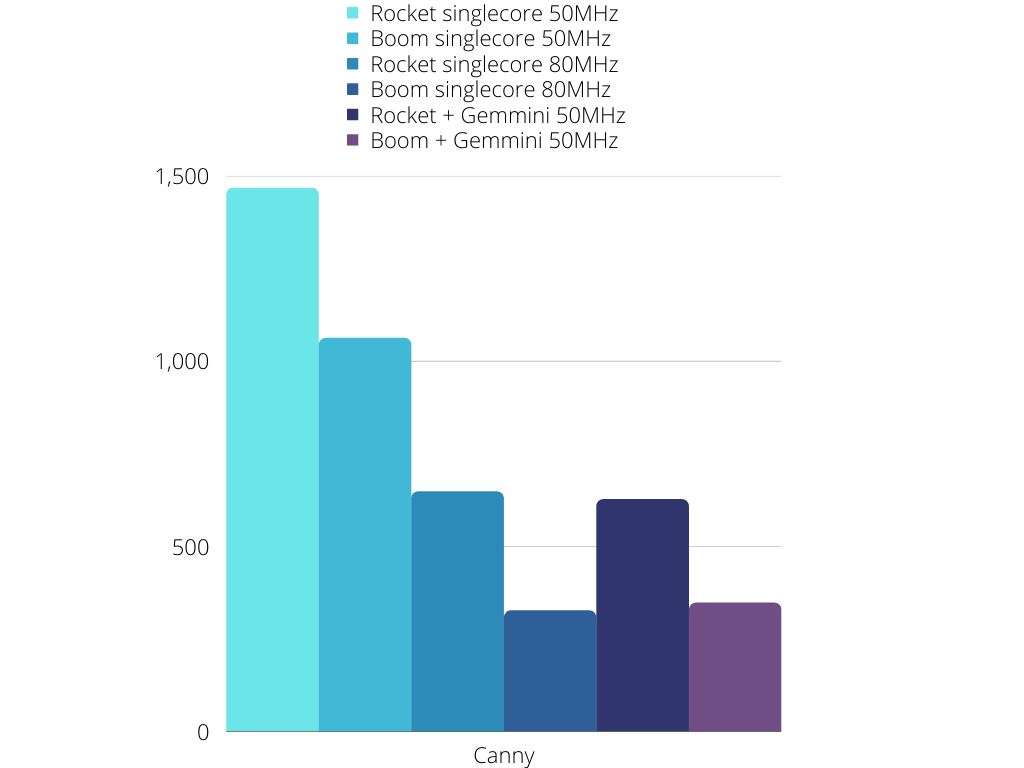}
\caption{\textcolor{black}{Time results when executing Workload 2 on Platforms 1 and 4 at 50MHz and 80MHz, and Workload 3 on Platforms 3 and 6 at 50MHz.}\label{fig:barchart_exp3}}
\end{figure*}

\textcolor{black}{Furthermore, in the graph shown in Fig.~\ref{fig:barchart_exp3} we can see the time corresponding to all the single core and heterogeneous experiments. The first thing we notice is that the out-of-order execution of the Boom core is beneficial for the Canny algorithm, leaving the Rocket core as the slowest by far at both 50 and 80MHz. Furthermore, we see how the combination of the cores with the Gemmini accelerator at 50MHz gives us a similar time to the same cores without accelerator at 80MHz, which gives us a great benefit in terms of consumption by running at a lower clock frequency which should be taken into account in the field of autonomous vehicles, as it would provide greater autonomy. In addition, we note that the shortest time is under half a second, in particular 300ms, and we achieve it with the combination of the Boom core and the Gemmini accelerator at a clock frequency of 50MHz. Thus, a vehicle travelling at 50km/h could run the algorithm every 4 metres approximately and if necessary, options such as mounting several systems in parallel or slightly increasing the clock frequency for faster processing could be explored.}

In conclusion, for this application with small matrices, both platforms based on the BOOM core deliver similar performance (speedup of around \textcolor{black}{3.7x} with respect to the Rocket baseline), being the BOOM single core at 80MHz slightly faster than the BOOM + Gemmini at 50MHz. Even in this non favourable scenario, the accelerator allows to report high performance working at a lower frequency, being more power efficient than the single core platform running at higher frequency.

\section{Conclusions and future work}

\textcolor{black}{In this paper we have explored the acceleration of a line detection algorithm in the autonomous car environment using a heterogeneous system consisting of a general-purpose RISC-V core and a domain-specific accelerator. In particular, we analyzed the application to identify the most computationally intensive parts of the code and adapted it accordingly for more efficient processing.}

The first conclusion we extract from this work is that RISC-V architecture provides a hw-sw ecosystem that is well suited for IoT in general and autonomous vehicle systems in particular, due to its versatility and modularity, which allows to generate platforms adapted to different scenarios. In fact, in this work, we designed six different platforms covering a wide spectrum of alternatives: on one side single and dual core homogeneous systems, and on the other side heterogeneous platforms with a single core plus a matrix multiplication accelerator --all of them including high performance BOOM cores or more efficient Rocket cores.

Also, a multithreaded application with high data parallelism has been designed to analyze the performance of the homogeneous platforms built. Thus, it has been verified that multithreaded applications are being correctly simulated in the multicore platforms, achieving the expected speedup. Furthermore, the comparison between both types of cores determined that a single BOOM core is up to 2.19 times faster than a Rocket one.

Finally, the original application of line detection has been modified in order to decrease its execution time without losing accuracy, and it has also been adapted for bare metal and Gemmini execution. We simulated the application on all designed platforms. BOOM-based platforms reported the best performance numbers, achieving speedups of \textcolor{black}{3.7x} with respect to the baseline (a single Rocket core running at 50MHz), and being the single BOOM core running at 80MHz slightly faster than the BOOM + Gemmini platform at 50MHz. As previously stated, even working at a lower frequency the accelerator allows to report high performance, being more power efficient than the single core counterpart working at a higher frequency. \textcolor{black}{It is worth noting that our goal in this work was to explore how an domain-specific accelerator was able to accelerate the baseline execution (just using a conventional single core) in applications belonging to autonomous vehicles environment}.

As future work, other applications which involve multiplication of big matrices can be adapted to heterogeneous platforms in order to implement more of the functionalities required for autonomous vehicles. Moreover, Gemmini is expected to achieve much higher speedups for inference using neural networks, as shown in \cite{gemmini-dac21}, so exploring this issue constitutes an interesting avenue for future work.

 \competinginterests{
The authors have declared that no competing interests exist.
 }

 \funding{
The present work has been funded by the Comunidad de Madrid through project S2018/TCS-4423 and by the Ministry of Science, Innovation and Universities through project RTI2018-093684-B-I00.}

\authorcontributions{

MJB wrote the programs, conducted the experiments, analyzed the results and wrote the manuscript; KO and FC conceived the idea, analyzed the results and wrote the manuscript; FT revised the manuscript. All authors read and approved the final manuscript.
}

\begin{small}
\bibliography{references}
\end{small}


\cornersize{.2} 
\setlength{\fboxsep}{8pt}

\ovalbox {
\begin{minipage}{7cm}
\textbf{Citation:} M.J. Belda, K. Olcoz, F. Castro and F. Tirado. \emph{Optimization of a line detection algorithm for autonomous vehicles on a RISC-V with accelerator}. Journal of Computer Science \& Technology, vol. xx, no. x, pp. x–x, 202x.

\textbf{DOI:} 10.24215/16666038.18.e01

\textbf{Received:} August x, 2022 \textbf{Accepted:} xxx.

\textbf{Copyright:} This article is distributed under the terms of the Creative Commons License CC-BY-NC.
\end{minipage}
}

\end{document}